\documentclass[11pt,a4paper]{amsart}
\usepackage{amsmath, amssymb, amsfonts, amsthm}
\usepackage{latexsym}
\usepackage{amsmath}
\usepackage{amssymb}
\usepackage{amsthm}
\usepackage{amsxtra}
\usepackage{amsfonts}
\usepackage{mathrsfs}
\usepackage{amsbsy}
\usepackage{graphicx}

\usepackage{hyperref}

\newtheorem{teo}{Theorem}

\newtheorem{lem}{Lemma}[section]

\newtheorem{prop}{Proposition}[section]
\newtheorem{remark}{Remark}[section]

\numberwithin{equation}{section}

\newenvironment{dem}{\vspace{.2cm}\noindent {\bf Proof }\\}{\newline \hspace{1cm} \hspace{1cm}\flushright \hfill $\square$ \newline}

\newenvironment{demteo}[1]{\vspace{.2cm}\noindent {\bf Proof {#1}}\\}{\newline \hspace{1cm} \hspace{1cm}\flushright \hfill $\square$ \newline}

\renewcommand{\Re}{\operatorname{Re}\,}

\newcommand{\KK}{\mathcal K}

\newcommand{\n}{\noindent}

\newcommand{\ve}{\varepsilon}
\newcommand{\erre}{\mathbb{R}} 

\newcommand{\enne}{\mathbb{N} }

\newcommand{\de}{\delta} 
\newcommand{\al}{\alpha}

\newcommand{\f}{\frac}
\newcommand{\ba}{\begin{eqnarray}} \newcommand{\ea}{\end{eqnarray}}
\newcommand{\be}{\begin{equation}} \newcommand{\ee}{\end{equation}}
\newcommand{\bdm}{\begin{displaymath}} \newcommand{\edm}{\end{displaymath}} 
\newcommand{\brr}{\begin{array}}\newcommand{\err}{\end{array}}

\newcommand{\lf}{\left}
\newcommand{\ri}{\right}

\newcommand{\bml}{\begin{gather}} 
\newcommand{\eml}{\end{gather}}

\DeclareMathOperator{\sign}{sign}

\newcommand{\beq}{\begin{equation}}
\newcommand{\eeq}{\end{equation}}
\newcommand{\RE}{\mathbb{R}}

\renewcommand{\leq}{\leqslant}
\renewcommand{\geq}{\geqslant}
\renewcommand{\epsilon}{\varepsilon}

\title[]{The NLS equation in dimension one with spatially concentrated nonlinearities: 
the pointlike limit}

\author[]{Claudio Cacciapuoti}
\address{Hausdorff Center for Mathematics,
 Institut f\"ur Angewandte Mathematik, Endenicher Allee, 60, 53115 Bonn, Germany}
\email{cacciapuoti@him.uni-bonn.de}

\author[]{Domenico Finco}
\address{Facolt\`a di Ingegneria, Universit\`a Telematica
Internazionale Uninettuno,  Corso Vittorio Emanuele II, 39, 00186 Roma, Italy}
\email{d.finco@uninettunouniversity.net}

\author[]{Diego Noja}
\address{Dipartimento di Matematica e Applicazioni, Universit\`a
 di Milano Bicocca,  via Roberto Cozzi, 53, 20125 Milano, Italy}
\email{diego.noja@unimib.it} 

\author[]{Alessandro Teta}
\address{Dipartimento di Matematica G. Castelnuovo, Sapienza Universit\`a di Roma,  Piazzale  Aldo Moro, 5, 00185 Roma, Italy}
\email{teta@mat.uniroma1.it}
\date{}

\thanks{
D.F. and D.N.  acknowledge the support of FIRB 2012 project ``Dispersive dynamics: Fourier Analysis and Variational Methods''. C.C. acknowledges the support of the FIR 2013 project ``Condensed Matter in Mathematical Physics'' (code RBFR13NAET)}

\begin{document}
\begin{abstract}
In the present paper we study the following scaled nonlinear Schr\"odinger equation (NLS) in one space dimension:
$$
i\frac{d}{dt} \psi^{\varepsilon}(t) =-\Delta\psi^{\varepsilon}(t) + \frac{1}{\epsilon}V\left(\frac{x}{\epsilon}\right)|\psi^{\varepsilon}(t)|^{2\mu}\psi^{\varepsilon}(t) \quad \quad \epsilon>0\ ,\quad V\in L^1(\erre,(1+|x|)dx) \cap L^\infty(\erre) \ . $$
This equation represents a nonlinear Schr\"odinger equation with a spatially concentrated nonlinearity.
We show that in the limit $\epsilon\to 0$, the weak (integral) dynamics converges in $H^1(\erre)$ to the weak dynamics of the NLS with point-concentrated nonlinearity: 
$$
i\frac{d}{dt} \psi(t) =H_{\alpha}\psi(t)  . 
$$
where $H_{\alpha}$ is the laplacian with the nonlinear boundary condition at the origin $\psi'(t,0+)-\psi'(t,0-)=\alpha|\psi(t,0)|^{2\mu}\psi(t,0)$ and $\alpha=\int_{\erre}Vdx$. The convergence occurs for every $\mu\in \erre^+$ if $V \geq 0$ and for every $\mu\in (0,1)$ otherwise.
The same result holds true for a nonlinearity with an arbitrary number $N$ of 
concentration points.
\end{abstract}
\maketitle
\section{Introduction and result}
The nonlinear Schr\"odinger equation with spatially dependent nonlinearities has been considered in many contexts, both for its mathematical interest and for its relevance in several physical models. They represent situations in which there is a spatial inhomogeneity of the response of the medium to the wavefunction propagation. Here we are interested in the special case in which the nonlinearity is strongly concentrated in space around a finite number of points. More precisely, we investigate  the limit where    
 the nonlinearity becomes {\it strictly pointlike}. Such nonlinearities appear  in several physical applications,  for example:  nonlinear diffraction of electrons from a thin layer (see \cite{BKB}),    analysis of nonlinear resonant tunneling (see \cite{JPS}),   models of pattern formation and bifurcation of solitons in nonlinear media (see \cite{DM,KKK,LMF,MA,SKBRC} and references  therein). Rigorous analysis of asymptotic stability of standing waves for some of the above models can be found in \cite{BKKS,KKS}.
To make precise the considered problem, we will consider the following nonlinear Schr\"odinger type equation
\be\label{regolarizzata}
i\frac{d}{dt} \psi^{\varepsilon}(t) =-\Delta\psi^{\varepsilon}(t) + \frac{1}{\epsilon}V\left(\frac{x}{\epsilon}\right)|\psi^{\varepsilon}(t)|^{2\mu}\psi^{\varepsilon}(t)\ , \quad \quad \epsilon>0\ ,\quad V\in L^1(\erre,(1+|x|)dx) \cap L^\infty(\erre)  \ . 
\ee
When $\sign\big( \frac{1}{\epsilon}V(\frac{x}{\epsilon})\big)=\pm 1$ the nonlinearity is called focusing or attractive ($-$) and respectively defocusing or repulsive ($+$). In general both behaviours are admitted.\par\noindent
It is well known that for a large class of inhomogeneities $V$ the previous equation enjoys global well posedness in energy space $H^1(\erre)$ for every initial datum $\psi^{\ve}(0)=\psi_0$. Here by $H^1$-solutions we mean the solutions of the integral equation
\beq \label{regolarizzataint}
\psi^{\ve}(t,x)= U(t)\psi^{\varepsilon}(0, x) -{i} \int_0^t \!ds \int dy\; U(t-s,x-y)\, \frac{1}{\epsilon}V\left(\frac{y}{\epsilon}\right) \, |\psi^{\ve} (s,y) |^{2\mu} \psi^{\ve}(s,y) \qquad \psi^{\varepsilon}(0,\cdot):=\psi_0 \in H^1(\erre)
\eeq
where $U(t)$ is the free Schr\"odinger unitary group, whose convolution kernel is  
\be\label{group}U(t,x)= \f{1}{\sqrt{4\pi \, i\,  t} } e^{ i \f{x^2}{4t} }\ .\ee\par\noindent
For example, in the above hypotheses ($V\in L^1(\erre,(1+|x|)dx) \cap L^\infty(\erre) $), Corollary 6.1.2 of \cite{Caz03} applies and one has global existence of strong $H^1$-solutions for every initial datum $\psi_0$ in $H^1$ if $V\geq 0$ and $\mu> 0$ (defocusing case), and for $\mu<2$ if $V$ is negative in some open interval, in particular when it is everywhere nonpositive (focusing case). For $\mu=2$, the critical case, one has global existence for small data (see Remark 6.1.3 of \cite{Caz03}).\par\noindent   
We are interested in identifying the limit dynamics of the previous equation when $\epsilon\to 0$.
In view of the fact that $\lim_{\epsilon\to 0}\frac{1}{\epsilon}V(\frac{x}{\epsilon})=\alpha\delta_0$, where $\alpha=\int_{\erre} V  dx$, the natural candidate can be formally written as
 
\be{\label{limite}}
i\frac{d}{dt} \psi(t) =-\Delta\psi(t) + \alpha\delta_{0}|\psi(t)|^{2\mu}\psi(t)\ ,
\ee
where a meaning has to be given to the nonlinear term at the right hand side. To this end, we recall that a theory of NLS with a pointlike nonlinearity has been developed in several papers. For the present one dimensional case, the definition of the model and the global well posedness has been given in {\cite {AT}}. 
In {\cite {KK}} a second proof of global well posedness is given, making use of a nonlocal approximation of pointlike nonlinearities which is related to the topic treated in the present paper and that will be commented upon later. 
The definition of a pointlike nonlinearity is mimicked on the case of a {\it linear} delta interaction, often called delta potential ({\cite {Albeverio}} and references therein). In the one dimensional case a delta interaction of strength $\alpha $ at $x=0$ is the singularly perturbed laplacian $(D(-\Delta_{\alpha}), -\Delta_{\alpha})$ given as follows

\be\label{dom1}
D (-\Delta_\alpha) \  : = \ \left\{ \phi \in H^2 (\erre \backslash \{0 \})\cap H^1 (\erre), 
\ \phi' (0+) - \phi' (0-) =  \alpha \phi (0) \right\}\, \quad {\rm and}\quad  -\Delta_{\alpha} \phi \  = - \phi^{\prime \prime}\ ,\ \ x\neq 0 
\ee
The corresponding quadratic form  $(D(Q_{\alpha}), Q_{\alpha})$ is given by
\be \label{quadratic}
D(Q_{\alpha})=H^1({\erre}) \ \ {\rm and} \ \ Q_{\alpha}(\phi)=\int_{\erre} |\phi'(x)|^2\ dx + \alpha|\phi(0)|^2\ .
\ee
It is immediate to extend this definition to the case of $N$ singularity points $y_k$, $k=1,\ldots,N$.   
\par\noindent
The idea is to modify the boundary condition and to consider the strength $\alpha$ depending on the wavefunction itself. 
\par\noindent
Following the cited literature, we take as a definition of a nonlinear interaction (of power type) concentrated at the point $x=0$ the nonlinear operator $(D(H_{\alpha,\mu}), H_{\alpha,\mu})$ given by 
\be\label{dom2}
D (H_{\alpha,\mu}) \  : = \ \left\{ \phi \in H^2 (\erre \backslash \{0 \})\cap H^1 (\erre), 
\ \phi' (0+) - \phi' (0-) =  \alpha(\phi) \phi (0) \right\}\, \quad {\rm and}\quad  H_{\alpha,\mu} \phi \  = -\phi^{\prime \prime}\ ,\ \ x\neq 0 
\ee
where
\be\label{nlboundary}
\alpha(\phi)=\alpha|\phi(0)|^{2\mu}\ ,\ \  \alpha \in \erre\ . 
\ee
\par\noindent
Obviously the boundary condition reduces to the linear one when $\mu=0$.\par\noindent
In \cite{AT} it is shown that the time dependent nonlinear Schr\"odinger equation with nonlinear generator given by $H_{\alpha,\mu}$,
\be \label{concentrated}
i\frac{d}{dt}\psi(t)= H_{\alpha,\mu}\psi(t)\ ,
\ee
is globally well posed for every initial datum $\psi(0) \in H^1(\erre)$ for $\mu>0$ if $\alpha>0$ (defocusing concentrated nonlinearity) and for $\mu\in(0,1)$ if $\alpha<0$ (focusing concentrated nonlinearity). By this we mean (see \cite{AT} for details) that there exists a unique global $H^1$-solution of the integral equation
\beq \label{concentratedint}
\psi(t,x)= U(t)\psi_0 -i\al \int_0^t ds\; U(t-s, x)\, q (s) \qquad  q (s)= |\psi (s,0) |^{2\mu} \psi(s,0)  \qquad \psi(0,\cdot)=\psi_0 \in H^1(\erre)
\eeq
\par\noindent
Notice, in the nonrepulsive case, the loss in the range of admissible exponents $\mu$ in passing from regular to pointlike spatial dependence.\par\noindent
More generally one can consider a nonlinearity with a finite number $N$ of {\it distinct} concentration points at, say, $y_1,\ldots,y_N$ and corresponding spatial modulations $V_k$. We can admit arbitrary signs for  the means $\int_{\erre} V_k \ dx$ and different powers for the nonlinearities. \par\noindent
The regularized equation in integral form is then
\beq \label{regolarizzataint2}
\psi^{\ve}(t,x)= U(t)\psi_0 -{i} \sum_{k=1}^N \int_0^t \!\!ds\! \int \!dy\; U(t-s,x-y)\, \frac{1}{\epsilon}V_{k}\!\left(\frac{y-y_k}{\epsilon}\!\right)  |\psi^{\ve} (s,y) |^{2\mu_k} \psi^{\ve}(s,y) \qquad \psi^{\ve}(0,\cdot):=\psi_0 \in H^1(\erre)
\eeq
while the limit equation is 
\beq\label{concentratedint2}
\psi(t,x)= U(t)\psi_0 -i\sum_{k=1}^N  \al_k \int_0^t ds\; U(t-s, x)\, q_k (s) \qquad  q_k (s)= |\psi (s,y_k) |^{2\mu_k} \psi(s,y_k)  \qquad \psi(0,\cdot):=\psi_0 \in H^1(\erre)
\eeq 
For this equation (see \cite{AT}) global well posedness  is known to hold under the hypotheses $\mu_k\in (0,1)$ if at least one $\alpha_k$ is negative, and for $\mu_k>0$ otherwise.
The main result of the paper is the following
\begin{teo}\label{teokappa}
For every $k=1,\ldots,N$, take $ V_k\in L^1(\erre,(1+|x|)dx) \cap L^\infty(\erre)$.  Let $V_k\geq 0$ or $\mu_k\in (0,1)$ and let  $\psi^{\varepsilon}(t)$ and $\psi(t)$ be the $H^1$-solutions of \eqref{regolarizzataint2} and \eqref{concentratedint2} with initial data $\psi^{\varepsilon}(0)=\psi(0)\equiv \psi_0$ and 
$
\alpha_k=\int V_k \ dx$. 
Then for any $ T\in \erre^+$ one has
\beq
\sup_{t\in [0,T]}\|\psi^{\ve} (t)-\psi(t)\|_{H^1}\to 0\quad\textrm{as }\;\; \ve\to 0\ .
\eeq
\end{teo}
\noindent
Section \ref{s:2} contains the proof of Theorem \ref{teokappa}. Several remarks to our main result are postponed at the end of the section. 

\noindent
In what follows $c$ denotes a generic positive constant which does not depend on $\ve$ and $t$ (but may depend on $T$, $\psi_0$ and $V_k$) and whose value may change form line to line.\par\noindent
Moreover we will denote by $||\cdot||$ the $L^2$-norm and with $||\cdot||_p$ the $L^p$-norm.

\section{Pointlike limit: convergence of spatially regular dynamics and proof of Theorem \ref{teokappa}}
\label{s:2}
As explained in the introduction, we compare two evolution problems. The first one, which we will call the rescaled problem is given by \eqref{regolarizzataint2} which is here rewritten as
\beq \label{rescaled}
\psi^{\ve}(t,x)= U(t)\psi_0(x) -\f{i}{\ve} \sum_{k=1}^N \int_0^t ds \int dy\; U(t-s,x-y)\, V^{\ve}_k(y-y_k) \, |\psi^{\ve} (s,y) |^{2\mu_k} \psi^{\ve}(s,y) \qquad \psi_0 \in H^1(\erre)
\eeq
where (notice the definition of $V^{\varepsilon}_k$)
\be
V^{\ve}_k (x) = V_k \lf( \f{x}{\ve} \ri)\ ,\quad \quad V_k\in L^1(\erre,(1+|x|)dx) \cap L^\infty(\erre), \quad\quad k=1,\ldots,N\ .
\eeq
It is a standard matter to show that the $L^2$-norm (``mass'') of the solution is conserved along the flow, and that it admits a conserved energy
\beq \label{rescaledenergy}
E^{\ve}(\psi^{\varepsilon}(t)) = \int dx\; |{\psi^{\ve}}'  (t,x) |^2 + \sum_{k=1}^N  \f{1}{\mu_k+1}  \int dx\; \f{1}{\ve} \, V^{\ve}_k (x-y_k)  |\psi^{\ve}(t,x)|^{2\mu_k+2}
\eeq
\par\noindent
We want to compare problem \eqref{rescaled} with the second evolution problem, the limit problem given by \eqref{concentratedint2}
with
$
\al_k = \int  V_k(x) \ dx.
$
\par\noindent
In \cite{AT} it is shown that the limit problem conserves mass and energy, where now the energy functional is defined by
\beq \label{limitenergy}
E(\psi(t)) = \int dx\; |\psi ' (t,x) |^2 +\sum_{k=1}^N   \al_k\, \f{1}{\mu_k+1} |\psi(t,y_k)|^{2\mu_k +2} 
\eeq
As recalled in the introduction, the solution $\psi(t)$ is global if $\al_k>0$ and  $\mu_k>0$ or $\al_k<0$ and $0<\mu_k<1$. 

\noindent
A main preliminary result used in the sequel is the following
\begin{prop}\label{uniformbound}
Take $V_k\in L^1(\RE)\cap L^{\infty}(\RE)$. Assume that $V_k\geq 0$ or $0< \mu_k< 1$ in the rescaled problem. Then the solution $\psi^{\ve}(t) $ of \eqref{regolarizzataint2} satisfies the following a priori estimate uniformly in $\ve$ and in $t$:
\beq \label{aprio1}
\sup_{t\in \erre}\| \psi^{\ve}(t) \|_{H^1} \leq c\ .
\eeq
\end{prop}
\begin{dem}
Due to mass conservation  it is sufficient to prove that
\[
\int dx \; |{{\psi^{\ve}}'} (t,x) |^2 \leq c
\]
We rewrite the energy in the following way
\[
E^{\ve}(\psi^{\ve}(t)) = \int dx\; |{\psi^{\ve}}' (t,x) |^2 + \sum_{k=1}^N  \f{1}{\mu_k+1}  \int dx\;  \, V_{k} (x)  |\psi^{\ve}(t,\ve x+y_k)|^{2\mu_k+2}
\]
For any $k=1,...,N$ we  rewrite $V_k=V_{k+}-V_{k-}$, where $V_{k+}\geq0$ and $V_{k-}\geq0$ are respectively  the positive and negative part of $V_k$. Let $\KK \subset \{1, \ldots , N\}$ be the set of indices such that $V_k$ is not positive. By energy conservation, Gagliardo-Nirenberg inequality  and neglecting the
terms such that $k\in \KK^c$, we have
\begin{align*}
E^\ve (\psi_0) &= E^\ve (\psi^{\ve} (t) )  \\
&\geq \int dx\; |{\psi^{\ve}}' (t,x) |^2 - \sum_{k\in \KK}  \f{1}{\mu_k+1}  \int dx\;  \, V_{k-} (x)  |\psi^{\ve}(t,\ve x+y_k)|^{2\mu_k+2} \\
& \geq  \int dx\; |{\psi^{\ve}}' (t,x) |^2 -\sum_{k\in \KK}  \f{1}{\mu_k+1} \|\psi^{\ve}(t)\|_\infty^{2\mu_k+2}  \int dx\;  \, V_{k-} (x)    \\
& \geq \| {\psi^{\ve}}' (t)\|^2 -c  \sum_{k\in \KK}   \| {\psi^{\ve}}' (t)\|^{1+\mu_k}
\end{align*}
Since
\[
\sup_{\varepsilon\in[0,1]} E^\ve(\psi_0) \leq \|\psi_0'\|^2 + \|\psi_0\|_\infty^{2\mu+2} \sum_{k=1}^N \frac{1}{\mu_k+1} \int dx \; \,V_{k+}(x) \equiv K
\]
 we finally get the inequality
\[
\| {\psi^{\ve}}' (t)\|^2 -c \sum_{k\in \KK}   \| {\psi^{\ve}}' (t)\|^{1+\mu_k}\leq K
\]
which implies $ \| {\psi^{\ve}}' (t)\| \leq c$ since,  for $k \in \KK$,   $\mu_k<1$ holds true.
\end{dem}

\n
Notice that the previous proposition implies an a priori bound for the $L^\infty$-norm of $\psi^{\ve} (t)$ uniformly in $\ve$ 
and $t$, that is
\beq \label{aprio2}
|\psi^{\ve} (t,x)|\leq c\ .
\eeq
We recall that also $\psi(t)$ enjoys a similar bound.

\n
Now we consider the convergence at the nonlinear defects locations $y_k,\ k=1,\cdots,N$.
\begin{lem} \label{lemma}
Take $ V_k\in L^1(\erre,(1+|x|)dx) \cap L^\infty(\erre)$.  Assume that $V_k\geq 0$ or $0<\mu_k< 1$ and $\delta\in (0,\frac{1}{2})$. Then for any $T \in \erre^+$ 
we have 
\beq \label{punto0}
\sup_{t\in [0,T]}|\psi^{\ve} (t,y_j)-\psi(t,y_j)|\leq c\ {\ve}^\de \qquad j=1,\ldots, N
\eeq
\end{lem}
\begin{dem}
Fix $j$ and let us rewrite \eqref{rescaled} as 
\[
\psi^{\ve}(t,x)= U(t)\psi_0(x) -{i}\sum_{k=1}^N \int_0^t ds\;\int dy\;  U(t-s,\, x-y_k -\ve y)\, V_k(y)\, |\psi^{\ve} (s,\ve y+y_k) |^{2\mu_k} \psi^{\ve}(s,\ve y+y_k) 
\]
which gives for $x=y_j$
\[
\psi^{\ve}(t,y_j)= U(t)\psi_0(y_j) -{i}\sum_{k=1}^N \int_0^t ds\;\int dy\;  U(t-s,\, y_j-y_k -\ve y)\, V_k(y)\, |\psi^{\ve} (s,\ve y+y_k) |^{2\mu_k} \psi^{\ve}(s,\ve y+y_k)
\]
to be compared with
\[
\psi(t,y_j)= U(t)\psi_0(y_j) -i \sum_{k=1}^N \al_k \int_0^t ds\; U(t-s, y_j-y_k)\, |\psi (s,y_k) |^{2\mu_k} \psi(s,y_k)
\]
Adding and subtracting suitable terms, we have the following identity
\[
\psi^{\ve}(t,y_j)-\psi(t,y_j) =  -i  \sum_{k=1}^N \al_k \int_0^t ds\; U(t-s, y_j-y_k)\,\lf( |\psi^{\ve} (s,y_k) |^{2\mu_k} \psi^{\ve}(s,y_k)  - |\psi (s,y_k) |^{2\mu_k} \psi(s,y_k) \ri) +
{\mathcal R}^{\ve}(t)
\]
with a remainder given by
$$
{\mathcal R}^{\ve}(t)={\mathcal R}^{\ve}_1 (t)+ {\mathcal R}^{\ve}_2(t)\ ,$$
\begin{align*}
{\mathcal R}^{\ve}_1 (t)& = -{i}\sum_{k=1}^N  \int_0^t ds\;\int dy\; \big[ U(t-s,\,y_j-y_k- \ve y)- U(t-s\,,\,y_j-y_k )\big] V_k (y)\, |\psi^{\ve} (s,y_k+ \ve y) |^{2\mu_k} \psi^{\ve}(s,y_k+\ve y) \\
 {\mathcal R}^{\ve}_2(t)& =  -{i}\sum_{k=1}^N  \int_0^t ds\; U(t-s,\, y_j-y_k )\int dy\; V_k(y)\,
\lf( |\psi^{\ve} (s,y_k+\ve y) |^{2\mu_k} \psi^{\ve}(s,y_k + \ve y) - |\psi^{\ve} (s,y_k) |^{2\mu_k} \psi^{\ve}(s,y_k) \ri)
\end{align*}
We provide an estimate of the remainder terms using the previously shown a priori bounds and the following elementary estimate
\beq \label{stimetta}
\lf| e^{i z}-1 \ri| \leq c |z|^{\de} \qquad \de<1/2, \ \ \ z\in \erre\ .
\eeq
Using \eqref{aprio2} and \eqref{stimetta} we have
\begin{align*}
|{\mathcal R}^{\ve}_1 (t)|
& \leq \sum_{k=1}^N \int_0^t ds\;\int dy\; \lf| U(t-s,\, y_j-y_k-\ve y)- U(t-s,\, y_j-y_k )\ri|\, |V_k(y)|\, |\psi^{\ve} (s,y_k+ \ve y) |^{2\mu_k+1}  \\
&\leq c\sum_{k=1}^N \int_0^t ds\;\int dy\; \lf| U(t-s,\, y_j-y_k-\ve y)- U(t-s,\, y_j-y_k )\ri|\, |V_k(y)|\\
&\leq  c\,\ve^{\de} \sum_{k=1}^N \int_0^t ds\;\f{1}{(t-s)^{1/2+\de}}   \int dy\; |2y(y_j- y_k)-\ve y^2|^\de \, |V_k (y)| =   c\ \ve^{\de}
\end{align*}
Concerning $ {\mathcal R}^{\ve}_2(t)$, we have
\[
| {\mathcal R}^{\ve}_2(t)|\leq  \sum_{k=1}^N \int_0^t ds\; \f{1}{\sqrt{4\pi(t-s)}}  \int dy\; |V_k (y)|\,
\lf| |\psi^{\ve} (s,y_k +\ve y) |^{2\mu_k} \psi^{\ve}(s,y_k +\ve y) - |\psi^{\ve} (s,y_k) |^{2\mu_k} \psi^{\ve}(s,y_k) \ri|
\]
Notice that, as a consequence of  \eqref{aprio1}  $$\| |\psi^{\ve} (s) |^{2\mu_k} \psi^{\ve}(s) \|_{H^1}\leq c\ \| \psi^{\ve} (s) \|^{2\mu_k+1}_{H^1} \leq c\ .$$ \\
Being every $f\in H^1(\erre)$ in the H\"older space $C^{1/2} (\erre)$, the following estimate hold true
\beq \label{stimapunt}
|f(y_k+ \ve y)-f(y_k)| 
\leq  \sqrt{\ve |y| }\, \| f\|_{H^1}\ .
\eeq
Applying this estimate to $|\psi^{\ve} (s) |^{2\mu_k} \psi^{\ve}(s)$ and using again Cauchy-Schwarz inequality, we obtain
\beq \label{resto2}
| {\mathcal R}^{\ve}_2(t)|\leq c\sqrt{\ve}  \sum_{k=1}^N  \int_0^t ds\; \f{1}{\sqrt{t-s}}  \int dy\; \sqrt{|y|}\,| V_k(y)|  \, \| |\psi^{\ve} (s) |^{2\mu_k} \psi^{\ve}(s) \|_{H^1} \leq  c\sqrt{\ve} 
\eeq
which finally implies 
\[
| {\mathcal R}^{\ve}(t)|\leq c\ {\ve}^\de
\]
\n
Using the above remainder estimate we can prove \eqref{punto0}. As a first step we have
\begin{align}
& |\psi^{\ve}(t,y_j)-\psi(t,y_j)| \nonumber \\
&\quad  \leq c\sum_{k=1}^N |\al_k|   \int_0^t ds\; \f{1}{\sqrt{t-s} } \,\lf| |\psi^{\ve} (s,y_k) |^{2\mu_k} \psi^{\ve}(s,y_k)  - |\psi (s,y_k) |^{2\mu_k} \psi(s,y_k) \ri| + c\ {\ve}^\de \nonumber  \\
 & \quad\leq c \sum_{k=1}^N   \int_0^t ds\; \f{1}{\sqrt{t-s} } \lf(  |\psi^{\ve} (s,y_k) |^{2\mu_k}   +|\psi (s,y_k) |^{2\mu_k} \ri) \lf|  \psi^{\ve}(s,y_k)  -  \psi(s,y_k) \ri| + c\ {\ve}^\de \nonumber \\
 & \quad \leq c\sum_{k=1}^N    \int_0^t ds\; \f{1}{\sqrt{t-s} } \lf|  \psi^{\ve}(s,y_k)  -  \psi(s,y_k) \ri| + c\ {\ve}^\de \ . \label{preiter}
\end{align}
%

\noindent
Let us define
\[
X^{\ve}(t)= 
\sum_{j=1}^N  |\psi^{\ve}(t,y_j)-\psi(t,y_j)| 
\]
then from \eqref{preiter} we have
\beq \label{iter}
X^{\ve}(t)\leq c_1  \int_0^t ds\; \f{1}{\sqrt{t-s} } X^{\ve}(s) + c_2 {\ve}^\de
\eeq
where $c_1$ and $c_2$ depend on $N$ and $V_k$. Using a standard argument in the theory of Abel integral operators (see \cite{GV}, ch. 7), one concludes that
$$
\sup_{t\in [0,T]} X^{\ve}(t) \leq  c \, \ve^{\delta}.
$$
\end{dem}

\n
We prove the $L^2$-convergence of the rescaled flow to the limit flow.
\begin{teo}Take $ V_k\in L^1(\erre,(1+|x|)dx) \cap L^\infty(\erre)$. Assume that $V_k\geq 0$ or $0<\mu_k< 1\ ,$ and let $\delta\in(0,\frac{1}{2})$. Then for any $T \in \erre^+$ we have 
\beq
\sup_{t\in[0,T]}\|\psi^{\ve} (t)-\psi(t)\|\leq c\ \ve^{\de} 
\eeq
\end{teo}
\begin{dem}
We use a decomposition similar to the one exploited in the previous lemma. We rewrite the two problems in integral form
\[
\psi^{\ve}(t,x)= U(t)\psi_0(x) -{i}\sum_{k=1}^N \int_0^t \!\! ds \int dy\;  U(t-s,\, x-y_k-\ve y)\, V_k(y)\, |\psi^{\ve} (s,y_k+ \ve y) |^{2\mu_k} \psi^{\ve}(s,y_k+\ve y) 
\]
\[
\psi(t,x)= U(t)\psi_0(x) -i\sum_{k=1}^N \al_k \int_0^t \!\! ds\; U(t-s, x-y_k)\, |\psi (s,y_k) |^{2\mu_k} \psi(s,y_k)
\]
Adding and subtracting suitable terms, we have
\[
\psi^{\ve}(t,x)-\psi(t,x) =  -i\sum_{k=1}^N \al_k \! \int_0^t \!\!ds\; U(t-s, x-y_k)\,\lf( |\psi^{\ve} (s,y_k) |^{2\mu_k} \psi^{\ve}(s,y_k)  - |\psi (s,y_k) |^{2\mu_k} \psi(s,y_k) \ri) +
{\mathcal T}^{\ve}(t,x)
\]
with remainder given by
\[
{\mathcal T}^{\ve}(t,x)={\mathcal T}^\ve_1 (t,x)+ {\mathcal T}^\ve_2(t,x)
\]
\begin{align*}
{\mathcal T}^\ve_1 (t,x)& = -{i}\sum_{k=1}^N \int_0^t \!\! ds \int \! dy\; \lf[ U(t-s,\,x- y_k-\ve y)- U(t-s,\,x -y_k)\ri]
\, V_k(y)\, |\psi^{\ve} (s,y_k+\ve y) |^{2\mu_k} \psi^{\ve}(s,y_k+\ve y) \\
 {\mathcal T}^\ve_2(t,x)& =  -{i}\sum_{k=1}^N  \int_0^t \!\! ds \;U(t-s,\,x-y_k )\!\int \!\! dy\; V_k(y)\,
\lf( |\psi^{\ve} (s,y_k+\ve y) |^{2\mu_k} \psi^{\ve}(s,y_k+\ve y) - |\psi^{\ve} (s,y_k) |^{2\mu_k} \psi^{\ve}(s,y_k) \ri)
\end{align*}
For  the first remainder $\mathcal T^{\ve}_1 (t)$ we have 
\begin{multline*}
\| {\mathcal T}^\ve_1 (t) \|^2 \! \leq \! c  \sum_{k=1}^N \int \! \!dx \lf|  \int_0^t \!
\! ds \!\int \!\! dy  \big[ U(t\!-\!s,\,x\!-\! y_k\!-\!\ve y)- U(t\!-\!s,\,x\!-\!y_k )\big] V_k(y)\, |\psi^{\ve} (s,y_k\!+\! \ve y) |^{2\mu_k} \psi^{\ve}(s,y_k\!+\!\ve y)   \ri|^2
\end{multline*}
Such an integral can be estimated using group property of $U(t)$ and \eqref{stimetta} in the following way. For every $f\in L^1(\erre, (1+|x|)dx)\ ,$
\begin{align*}
& \int dx \lf|  \int_0^t ds\;\int dy\; \big[ U(t-s,\,x-y_k- \ve y)- U(t-s,\,x-y_k )\big]\, f(y) \ri|^2 \\
& =  \int_0^t ds\, ds' \;\int dy\, dy' \; \overline{f(y)} f(y') \big( U(s-s', \ve y - \ve y') -  U(s-s', \ve y ) - U(s-s',  \ve y') + U(s-s', 0)\big) \\
& \leq  \int_0^t ds\, ds' \;\int dy\, dy' \; |f(y)| |f(y')| \big( | U(s-s', \ve y - \ve y') -  U(s-s', \ve y )| + |U(s-s', 0) - U(s-s',  \ve y') | \big) \\
& \leq c\ \ve^{2\de}  \int_0^t ds\, ds' \f{1}{|s-s'|^{1/2+\de}} \;\int dy\, dy' \; |f(y)| |f(y')| \lf( |y|^{2\de} + | y' |^{2\de}\ri)
\end{align*}
Using \eqref{aprio2} and the estimate above, we have
\[
\| {\mathcal T}^\ve_1 (t) \| \leq c\ \ve^{\de}  \| \psi^{\ve} \|_{\infty}^{2\mu+1}\sum_{k=1}^N  \int dy |V_k(y) | \, (1+|y|^{2\de})  \leq  c\, \ve^\de
\]
Concerning the estimate of $ {\mathcal T}^\ve_2(t)$, we use the following estimate
\beq \label{stimal2}
\lf\| \int_0^t U(t-s,\cdot) f(s) \, ds \ri\| \leq c \sup_{s\in[0,t]}| f(s)|\
\eeq
and we obtain 
\begin{align*}
\|  {\mathcal T}^{\ve}_2(t)\|
& \leq c \sum_{k=1}^N \sup_{s\in[0,t]}\int dy\, |V_k(y)|\,\lf| |\psi^{\ve} (s,y_k+ \ve y) |^{2\mu_k} \psi^{\ve}(s,y_k +\ve y) - |\psi^{\ve} (s,y_k) |^{2\mu_k} \psi^{\ve}(s,y_k) \ri| \\
& \leq c\sqrt{\ve}    \sum_{k=1}^N  \int dy\; \sqrt{|y|}\,| V_k(y)|  \, \sup_{s\in[0,t]}\| |\psi^{\ve} (s) |^{2\mu_k} \psi^{\ve}(s) \|_{H^1}^2 \leq  c\sqrt{\ve}
\end{align*} 
Therefore we have
\[
\| {\mathcal T}^\ve (t) \| \leq  c\ \ve^{\de} 
\]
Finally we arrive at the main estimate. Using \eqref{stimal2} and \eqref{punto0} we have
\begin{align*}
\|  \psi^{\ve}(t)-\psi(t)\|  & \leq   \sum_{k=1}^N|\al_k| \lf\| \int_0^t ds\; U(t-s, \cdot-y_k)\,\lf( |\psi^{\ve} (s,y_k) |^{2\mu_k} \psi^{\ve}(s,y_k)  - |\psi (s,y_k) |^{2\mu_k} \psi(s,y_k) \ri) \ri\| + c\, \ve^{\de}  \\
& \leq c  \sum_{k=1}^N  \sup_{ s\in [0,  t]} \lf| |\psi^{\ve} (s,y_k) |^{2\mu_k} \psi^{\ve}(s,y_k)  - |\psi (s,y_k) |^{2\mu_k} \psi(s,y_k) \ri| \ + c\,\ve^\de \\
& \leq c  \sum_{k=1}^N \sup_{s \in [0, t]} \lf( |\psi^{\ve} (s,y_k) |^{2\mu_k} + |\psi (s,y_k) |^{2\mu_k} \ri)  |  \psi^{\ve}(s,y_k)  - \psi(s,y_k) | \ + c\, \ve^{\de}  \\
& \leq c\,  \ve^\de
\end{align*}
\end{dem}


\begin{demteo}{of Theorem \ref{teokappa}}
By \eqref{aprio1} there is a subsequence, still denoted in the same way,  such that $\psi^{\ve}(t) \rightharpoonup \phi(t)$ in $H^1(\erre)$.
Notice that $\psi^{\ve}(t) \to \psi(t) $ in $L^2(\erre)$ and then a.e. $\phi(t) =  \psi(t) $.
Moreover, since
\[
\| \psi^{\ve}(t) - \psi(t) \|_{H^1}^2 =\| \psi^{\ve}(t) \|_{H^1}^2 + \| \psi(t) \|_{H^1}^2- 2 \Re ( \psi^{\ve} (t) , \psi(t) )_{H^1}
\]
it is sufficient to prove that   
\[
\lim_{\ve\to0}\| \psi^{\ve}(t) \|_{H^1}^2 = \| \psi(t) \|_{H^1}^2\ .
\]
In fact, due to the previous theorem it is sufficient to prove 
\[
\lim_{\ve\to0} \int dx\; |{\psi^{\ve}} ' (x) |^2 =  \int dx\; |\psi ' (x) |^2\ .
\]
Using the conservation of the  energy, we write 
\begin{align*}
 \int dx\; |{\psi^{\ve}}' (t,x) |^2 &= E^\ve (\psi_0) -  \sum_{k=1}^N \f{1}{\mu_k+1}  \int dx\; \f{1}{\ve} \, V^{\ve}_k (x-y_k)  |\psi^{\ve}(t,x)|^{2\mu_k+2}\\
 \int dx\; |\psi ' (t,x) |^2 &= E (\psi_0) -   \sum_{k=1}^N \al_k\, \f{1}{\mu_k+1} |\psi(t,y_k)|^{2\mu_k +2} \ .
\end{align*}
Since 
\[
 \lim_{\ve\to 0}E^\ve (\psi_0) =  E (\psi_0) 
\]
and, by lemma \ref{lemma}, 
\[
\lim_{\ve\to0}|\psi^{\ve}(t,y_k)|^{2\mu_k+2} =  |\psi(t,y_k)|^{2\mu_k +2} ,
\]
we have that 
\begin{align*}
&\lim_{\ve\to 0} \int dx\; |{\psi^{\ve}}' (t,x) |^2 \\
=& 
 \int dx\; |\psi ' (t,x) |^2  - \sum_{k=1}^N \frac{1}{\mu_k+1}\lim_{\ve\to 0}  \lf( \int dx\; \, V_k (x)  |\psi^{\ve}(t,y_k+\ve x)|^{2\mu_k+2} -  \al_k\,  |\psi^{\ve}(t,y_k)|^{2\mu_k +2} \ri).
\end{align*}
By  \eqref{stimapunt}, we have 
\begin{align*}
&\lf| \int dx\; \, V_k (x)  |\psi^{\ve}(t,y_k+\ve x)|^{2\mu_k+2} -  \al_k\,  |\psi^{\ve}(t,y_k)|^{2\mu_k +2} \ri|   \\
\leq & \int dx\; \, |V_k (x)| \, \lf|    |\psi^{\ve}(t,y_k +\ve x)|^{2\mu_k+2} -   |\psi^{\ve}(t,y_k)|^{2\mu_k +2} \ri| \\
\leq&  c \sqrt\ve \int  dx\; \, \sqrt{|x|} \, |V_k (x)|  \, \left \|    |\psi^{\ve}(t)|^{2\mu_k+2} \right\|_{H^1}  \leq c\sqrt\ve
\end{align*}
and this ends the proof.
\end{demteo}
%
%

\n
We conclude with some remarks and discussion of possible further developments.

\begin{remark}
There is a loss in the admitted range of nonlinearity powers guaranteeing global existence going from the regularized problem to the limit problem in  presence of attractive interactions: $\mu_k<2$ in the regularized problem and $\mu_k<1$ in the limit problem. This fact does not appear in the limit process at any stage, if not in the need of the uniform $H^1$ bound on $\psi^{\ve}(t)$ given by Proposition \ref{uniformbound}. It could be interesting a more detailed comprehension of this phenomenon.
\end{remark}
\begin{remark}
As regards Proposition \ref{uniformbound}, let us  consider for simplicity the case of a single inhomogeneity $V^{\epsilon}$, with $\int_{\erre} V^{\epsilon} dx >0$ but  having a nontrivial negative part $V^{\epsilon}_-$. It would be worth to understand whether the limitation $0<\mu<1$ is optimal.
\end{remark}
\begin{remark}
As noticed in the introduction, in \cite{KK} a proof of global well posedness of the limit problem is given, seemingly unaware of the already existing one. The paper is however interesting because in the course of the proof an approximating problem is constructed, corresponding to a way of recovering a delta potential from a regular interaction different from the one  exploited here. Namely, the approximating problem is the one corresponding to the nonlinear Schr\"odinger equation with a nonlocal interaction (``mean field''):
$$
i\frac{d}{dt}\psi^{\ve}(t)=-\Delta\psi^{\ve}(t) -\rho_{\epsilon}F(\langle \rho_{\epsilon},\psi^{\ve}(t)\rangle)
$$
where $\rho_{\epsilon}\rightharpoonup \delta_0$ in distributional sense and $F$ is the nonlinearity. The authors treat also nonlinearities different from powers, but in the focusing case they have to restrict to sublinear nonlinearities. 
It is a known fact from the theory of point interactions that if $F$ is the identity map the r.h.s. as a linear operator is a rank one perturbation of the laplacian converging in norm resolvent sense to the operator $\Delta_{\alpha}$ for a certain $\alpha$ (see \cite{Albeverio}). 
Well posedness for  the wave equation in one dimension and concentrated nonlinearities is studied along similar lines in \cite{Mart}.  
\end{remark}
\begin{remark}
Concentrated nonlinearities of power type in three dimensions are defined and studied in the NLS context in \cite{ADFT,ADFT2}. These models admit solutions of the kind of solitary (more precisely standing) waves. In \cite{Cecilia1} orbital and asymptotic stability of standing waves for NLS with concentrated nonlinearity is addressed. A definition of general concentrated nonlinearities with application to well posedness of the $1+3$ dimensional wave equation is given in \cite{NP}. Again in the wave case, it is easy to see that standing waves exist for the Klein-Gordon equation with concentrated nonlinearity. In view of this rich mathematical structure of dispersive PDE with  concentrated nonlinearities in dimension three, the extension of the result proved in the present paper to such a   case would be of   interest. 
We stress that, following the analogy with the linear case (see \cite{Albeverio}), the limit procedure in dimension three is more subtle and  requires a different  scaling from the one used in dimension one. 
We also mention that  essentially nothing is known, apart from the definition, concerning concentrated nonlinearities in two dimensions.  
\end{remark}
\begin{remark}
As regards the hypotheses on $V_k$, we notice the following. $V_k\in L^1(\erre)$ is of course a natural request, the only one needed in the linear case; $V_k\in L^\infty(\erre)$ is needed for having well posedness of the regularized problem (see the already quoted Corollary 6.2.1 in \cite{Caz03}); the condition $V_k\in L^1(\erre,(1+|x|)dx)$ is used in  Lemma 2.1 and Theorem 2, but, as it turns out from the proofs, it could be relaxed to  $V_k\in L^1(\erre, 1+|x|^\gamma)$ for a $\gamma\in (0,1)\ .$ Here we opted for the simplest statement.
\end{remark}

\medskip

\newcommand{\etalchar}[1]{$^{#1}$}
\providecommand{\bysame}{\leavevmode\hbox to3em{\hrulefill}\thinspace}
\providecommand{\MR}{\relax\ifhmode\unskip\space\fi MR }
\providecommand{\MRhref}[2]{%
  \href{http://www.ams.org/mathscinet-getitem?mr=#1}{#2}
}
\providecommand{\href}[2]{#2}

\vskip30pt
\end{document}